\documentstyle[epsf,amssymb,aps]{revtex}

\begin{document}
\newcommand{\mafigura}[4]{
  \begin{figure}[hbtp]
    \begin{center}
      \epsfxsize=#1 \leavevmode \epsffile{#2}
    \end{center}
    \caption{#3}
    \label{#4}
  \end{figure} }
\author{L. B. Leinson}
\address{Institute of Terrestrial Magnetism, Ionosphere and Radio Wave Propagation\\
RAS, 142092 Troitsk, Moscow Region, Russia }
\author{A. P\'{e}rez}
\address{Departamento de F\'{\i}sica Te\'{o}rica, Universidad de Valencia \\
46100 Burjassot (Valencia), Spain}
\draft
\title{Collective effects in $\nu \overline{\nu }$ synchrotron radiation from
neutron stars. }
\maketitle

\begin{abstract}
We have considered collective effects in $\nu \overline{\nu }$ synchrotron
radiation from an ultrarelativistic degenerate electron gas in neutron stars
with strong magnetic fields. For this problem we apply a calculation method
which explicitly makes use of the fact that the radiating electron moves
semi-classically, but takes into account the interaction among particles in
a quantum way. First we apply this method to calculate $\nu \overline{\nu }$
synchrotron radiation by an ultrarelativistic electron in vacuum and we
compare this result with that obtained previously by other techniques. When
a degenerate plasma is considered, we show that collective effects lead to
an essential enhancement (about three times) of the vector weak-current
contribution to neutrino pair emissivity.
\end{abstract}

\pacs{97.60.Jd,95.30.Cq,13.15.-f,52.25.Tx}

\section{Introduction}

Synchrotron radiation of neutrino pairs from a strongly magnetized
degenerate gas of ultrarelativistic electrons in a neutron star has been
studied by many authors (See \cite{Lands}-\cite{Kaminker3}). All cited
considerations of the problem are based on a single-particle approach, i.e.
except for the Pauli principle, no interaction among the electrons is taken
into account. However, due to the electromagnetic interaction, an electron
moving along a circular orbit involves in its motion a number of neighboring
electrons. In other words, the electron, moving in the electron gas, is
followed by a cloud of virtual particles and holes. When annihilating, they
also produce a flux of neutrino pairs going in the same direction as $\nu 
\overline{\nu }$ radiation of the initial electron. To evaluate this effect
at the lowest order of coupling constants one has to include in the matrix
element of weak interactions two Feynman diagrams, shown in Fig. 1.

By taking into account the Pauli principle, the first diagram describes the
mechanism of $\nu \bar{\nu}$ synchrotron radiation discussed in the
literature. The second diagram corresponds to the interaction of the initial
electron with a weak neutrino field via intermediate particle and hole
excitations, which are also included in the photon propagator. Naively, the
second diagram contribution to the matrix element might appear to be $%
e^{2}=1/137$ times smaller than the first one. However, as we will show,
this is not the case. In fact, the fine structure constant enters only in
the medium polarization function $\Pi \left( \omega ,{\bf k}\right) $, shown
in the diagram as a loop. So, the effective weak interaction via the medium
polarization introduces, as compared to the first diagram, an extra factor
which, in the case of a transversal virtual photon, is of the order $\Pi
/\left( q^{2}-\Pi \right) $ , where $q=\left( \omega ,{\bf k}\right) $ is
the total four-momentum carried out by the neutrino pair. To estimate this
factor one should take into account that synchrotron radiation from an
ultrarelativistic electron goes into a narrow cone with angle $\theta \sim
m/E$ relative to its velocity. Here $m$ and $E$ are the mass and the energy,
respectively, of the electron. By this reason one has (we set $\hbar
=c=k_{B}=1$) 
\begin{equation}
q^{2}\lesssim \frac{m^{2}}{E^{2}}\omega ^{2}\ll \omega ^{2}  \label{1I}
\end{equation}
i.e., the total four momentum transfer is much smaller than the total energy
of the radiated neutrino pair. Let us consider a degenerate electron gas
under the following condition \footnote{%
Note that in the opposite case $T\ll \omega _{0}$ , the population of
excited Landau levels above the Fermi energy exponentially tends to zero,
and the synchrotron radiation becomes not relevant as a mechanism of
neutrino pair emission.}: 
\begin{equation}
\frac{eB}{TE_{F}}\ll 1  \label{2I}
\end{equation}
Since $\omega $ is of the order of the medium temperature $T$, one has $%
\omega \gg \omega _{0}$ , where $\omega _{0}=eB/E_{F}$ is the Larmor
frequency of degenerate electrons near the Fermi energy $E_{F}$. Then one
can neglect the small contribution of the magnetic field to the medium
polarization, and use the polarization function $\Pi $ of an isotropic
electron gas which, when $\omega \rightarrow k$, can be estimated as the
square of the plasma frequency, defined as 
\begin{equation}
\omega _{p}^{2}=\frac{4}{3\pi }e^{2}E_{F}^{2}  \label{3I}
\end{equation}
Thus, one obtains 
\begin{equation}
\left| \frac{\Pi }{q^{2}-\Pi }\right| \sim \left| \frac{\omega _{p}^{2}}{%
q^{2}-\omega _{p}^{2}}\right| \sim 1  \label{4I}
\end{equation}
because 
\begin{equation}
\frac{q^{2}}{\omega _{p}^{2}}\sim \frac{1}{e^{2}}\frac{m^{2}}{E_{F}^{2}}%
\frac{T^{2}}{E_{F}^{2}}  \label{5I}
\end{equation}
and the last ratio is small for a degenerate ultrarelativistic electron gas.

Published calculations of $\nu \bar{\nu}$ emissivity due to synchrotron
radiation from an electron gas in the neutron star made use of exact wave
functions of the electron in a uniform magnetic field. In this case,
integration of the squared matrix element leads to either Laguerre or Bessel
functions with a complicated behavior, and summation over all initial and
final Landau states is a delicate problem. Actually, however, magnetic
fields inside neutron stars typically satisfy the condition $\omega _{0}\ll
E_{F}$ , which means that electrons near the Fermi surface move
semi-classically, and there is no reason to use exact Landau states in order
to calculate the neutrino pair emissivity. In order to study $\nu \bar{\nu}$
synchrotron radiation of ultrarelativistic electrons, we apply an
alternative method of calculation, which explicitly makes use of the fact
that the radiating particle moves semi-classically, although it takes into
account the interaction among particles in a quantum way. This method has
been developed by Baier and Katkov \cite{Baier} to consider $\gamma $
-synchrotron radiation in vacuum. A detailed derivation of this method
applied to $\gamma $-radiation is given, for instance, in \cite{Landau}.

This paper is organized as follows. In order to test our calculation
technic, in Section II we focus on $\nu \bar{\nu}$ synchrotron radiation of
ultrarelativistic electrons in strongly magnetized vacuum, i.e. we do not
set any limitation on the radiated energy of the neutrino pair. We show that
the total rate of $\nu \bar{\nu}$ decay, obtained by our calculation
technic, coincides with that obtained by other calculation methods \cite
{Baier66}-\cite{Ternov}. We also derive a simple formula for the
differential rate of decay in vacuum. When multiplied by the appropriate
statistical factors, thus taking into account the Pauli principle, this
formula can also be used for the case of an ultrarelativistic electron gas
in order to evaluate the contribution of the first diagram shown in Fig. 1,
as it is made in Section III. In Section IV we discuss some modifications to
be made in our calculation in order to take into account the second diagram
shown in Fig.1. We consider the medium polarization tensor, as well as the
photon propagator to be used for an ultrarelativistic degenerate electron
plasma. In Section V we calculate the contribution of collective effects to
the differential rate of neutrino pair emission. By adding these
contributions to that obtained in Section III we arrive to a very simple
analytic expression for the differential rate of $\nu \bar{\nu}$ synchrotron
radiation under the physical conditions we considered. In Section VI we
apply this formula to evaluate the energy loss and compare our result with
that obtained without taking into account collective effects . We end in
Section VII with some discussion and conclusions.

\section{\protect\bigskip $\nu \bar{\nu}$ synchrotron radiation by electrons
in vacuum}

An ultra-relativistic particle radiates into a narrow cone with angle $%
\theta \sim m/E$ relative to its velocity. By this reason, in the process
under consideration, the transferred four-momentum is much smaller than the $%
Z$-boson mass 
\begin{equation}
q^2=\omega ^2-{\bf k}^2\sim \frac{\omega ^2}{\gamma ^2}<m^2\ll M_Z^2
\label{k2}
\end{equation}
Therefore one can use an effective current-current interaction : 
\begin{equation}
{\cal L=}\frac{G_F}{\sqrt{2}}\bar{\nu}\gamma _\mu \left( 1-\gamma _5\right)
\nu \bar{\psi}\gamma ^\mu \left( C_V-C_A\gamma _5\right) \psi  \label{L}
\end{equation}
where $\nu $ represents the neutrino field and $\psi $ stands for the
radiating particle field. The coefficients $C_V$ and $C_A$ are the vector
and axial-vector weak coupling constants, respectively, of the radiating
particle, which also depend on the neutrino type. In the case of radiating
electrons, the combinations that arise, by summing over all neutrino
species, are 
\begin{equation}
\sum_\nu C_V^2=\frac 34-2\sin ^2\vartheta _W+12\sin ^4\vartheta _W\simeq
0.911  \label{CV}
\end{equation}
\begin{equation}
\sum_\nu C_A^2=\frac 34  \label{CA}
\end{equation}

By modification of the method described in the introduction to weak
processes, one obtains the following formula for the differential decay
width : 
\begin{eqnarray}
d\Gamma &=&\frac{G_F^2}{8\omega _1\omega _2}\frac{d^3k_1}{\left( 2\pi
\right) ^3}\frac{d^3k_2}{\left( 2\pi \right) ^3}{\sf Tr}\left( \left( \gamma
k_2\right) \gamma _\mu \left( 1-\gamma _5\right) \left( \gamma k_1\right)
\gamma _\nu \left( 1-\gamma _5\right) \right) \times  \label{1} \\
&&\times \int_{-\infty }^\infty d\tau e^{-i\omega \tau }\left\langle i\left| 
\hat{Q}^{\mu \dagger} \left( \frac \tau 2\right) \hat{Q}^\nu \left( -\frac %
\tau 2\right) \right| i\right\rangle  \nonumber
\end{eqnarray}
where $k_1=\left( \omega _1,{\bf k}_1\right) $ and $k_2=\left( \omega _2,%
{\bf k}_2\right) $ are the neutrino and antineutrino four-momenta,
respectively, $|i>$ represents the initial quantum state of the electron and 
$\hat{Q}^\mu \left( t\right) $ denotes the following Heisenberg operator : 
\begin{equation}
\widehat{Q}^\mu \left( t\right) =\frac{\bar{u}_f\left( \widehat{p}\right) }{%
\left( 2\widehat{H}\right) ^{1/2}}\gamma ^\mu \left( C_V-C_A\gamma _5\right)
\exp \left[ -i{\bf k}\widehat{{\bf r}}\left( t\right) \right] \frac{%
u_i\left( \widehat{p}\right) }{\left( 2\widehat{H}\right) ^{1/2}}  \label{2}
\end{equation}
with $q=\left( \omega ,{\bf k}\right) $ being the total four momentum of the
neutrino pair $q=k_1+k_2$ and 
\begin{equation}
u\left( \widehat{p}\right) =\left( 
\begin{array}{c}
\left( \widehat{H}+m\right) ^{1/2}w \\ 
\left( \widehat{H}+m\right) ^{-1/2}\left( {\bf \sigma }\widehat{{\bf p}}%
\right) w
\end{array}
\right)  \label{3}
\end{equation}
the fourspinor operator representing the symbolic solution of the Dirac
equation for a electron with a Hamiltonian $\widehat{H}$ , which includes
the effect of the external field. It can be obtained from the plane wave
fourspinor solution by replacing ${\bf p}$ and $E$ to the operators $%
\widehat{{\bf p}}=\widehat{{\bf P}}-e{\bf A=-}i{\bf \nabla -}e{\bf A,}$ $%
\widehat{H}=\left( \widehat{{\bf p}}^2+m^2\right) ^{1/2}$. The spin state of
the electron is determined by the spinor $w$. Finally, $\widehat{{\bf r}}%
\left( t\right) $ stands for the position operator of the electron.

By using the following identity 
\begin{eqnarray}
&&\int \frac{d^{3}k_{1}}{2\omega _{1}}\frac{d^{3}k_{2}}{2\omega _{2}}\delta
^{\left( 4\right) }\left( q-k_{1}-k_{2}\right) {\sf Tr}\left( \left( \gamma
k_{2}\right) \gamma _{\mu }\left( 1-\gamma _{5}\right) \left( \gamma
k_{1}\right) \gamma _{\nu }\left( 1-\gamma _{5}\right) \right)  \nonumber \\
&=&\frac{4\pi }{3}\theta \left( q^{2}\right) \theta \left( \omega \right)
\left( q_{\mu }q_{\nu }-g_{\mu \nu }q^{2}\right)  \label{4}
\end{eqnarray}
one can reduce the number of integrations in Eq.(\ref{1}) 
\begin{eqnarray}
d\Gamma &=&\frac{4\pi }{3}\frac{G_{F}^{2}}{2}\frac{d^{4}k}{\left( 2\pi
\right) ^{6}}\theta \left( q^{2}\right) \theta \left( \omega \right) \left(
q_{\mu }q_{\nu }-g_{\mu \nu }q^{2}\right) \times  \label{5} \\
&&\times \int_{-\infty }^{\infty }d\tau e^{-i\omega \tau }\left\langle
i\left| \hat{Q}^{\mu \dagger }\left( \frac{\tau }{2}\right) \hat{Q}^{\nu
}\left( -\frac{\tau }{2}\right) \right| i\right\rangle  \nonumber
\end{eqnarray}
Since we assume the ultra-relativistic electron motion to be semi-classical,
in the expression for $\widehat{Q}^{\mu }\left( t\right) $ only the
non-commutativity of the electron operators with the neutrino field operator 
$\exp \left( -i{\bf k}\widehat{{\bf r}}\right) $ needs to be taken into
account. After this, we can replace the electron operators by their
classical values. Following \cite{Landau} we obtain: 
\begin{equation}
e^{-i\omega \tau }Q_{1}^{\mu \dagger }Q_{2}^{\nu }\thickapprox \exp \left[ i%
\frac{E}{E^{\prime }}\left( {\bf kr}_{2}{\bf -kr}_{1}-\omega \tau +\frac{%
\left( \omega ^{2}-{\bf k}^{2}\right) }{2E}\tau \right) \right] R_{2}^{*\mu
}R_{1}^{\nu }  \label{6}
\end{equation}
with 
\begin{equation}
R^{\mu }\left( t\right) =\frac{u_{f}^{\dagger }\left( p^{\prime }\right) }{%
\left( 2E^{\prime }\right) ^{1/2}}\gamma ^{0}\gamma ^{\mu }\left(
C_{V}-C_{A}\gamma _{5}\right) \frac{u_{i}\left( p\right) }{\left( 2E\right)
^{1/2}}  \label{7}
\end{equation}
here, $E^{\prime }=E-\omega $ and ${\bf p}^{\prime }\left( t\right) ={\bf p}%
\left( t\right) -{\bf k}$. We assume that the initial state of the electron
has a momentum ${\bf p}\left( t\right) =E{\bf v}\left( t\right) $. We note
that the last term in exponential (\ref{6}) appears because $\omega ^{2}-%
{\bf k}^{2}\geqslant 0$ for the radiated neutrino pair. Here and
henceforward, suffixes $1$ and $2$ denote the values of quantities at the
times $t=-\frac{1}{2}\tau $ and $t=+\frac{1}{2}\tau $, respectively.
Therefore, ${\bf r}_{1}$ and ${\bf r}_{2}$ are the electron coordinates at
corresponding times.

An explicit calculation gives : 
\begin{eqnarray}
R^{0} &=&C_{V}\frac{w_{f}\left[ 2E\left( E+m\right) -E\left( {\bf k\cdot v}%
\right) -iE\left( \left( {\bf k\times v}\right) \cdot {\bf \sigma }\right)
\right] w_{i}}{2\sqrt{EE^{\prime }}\sqrt{\left( E^{\prime }+m\right) }\sqrt{%
\left( E+m\right) }}+  \label{8} \\
&&+C_{A}\frac{w_{f}\left[ 2\left( E+m\right) E\left( {\bf \sigma \cdot v}%
\right) -\left( E+m\right) \left( {\bf \sigma \cdot k}\right) \right] w_{i}}{%
2\sqrt{EE^{\prime }}\sqrt{\left( E^{\prime }+m\right) }\sqrt{\left(
E+m\right) }}  \nonumber
\end{eqnarray}
\begin{eqnarray}
{\bf R} &=&C_{V}\frac{w_{f}\left[ \left( 2E+2m-\omega \right) E{\bf v}%
-i\omega E\left( {\bf v\times \sigma }\right) +i\left( E+m\right) \left( 
{\bf k\times \sigma }\right) \right] w_{i}}{2\sqrt{EE^{\prime }}\sqrt{\left(
E^{\prime }+m\right) }\sqrt{\left( E+m\right) }}+  \label{9} \\
&&C_{A}\frac{w_{f}\left[ \left( \left( 2m-\omega \right) \left( E+m\right)
+E\left( {\bf k\cdot v}\right) \right) {\bf \sigma }+2E^{2}\left( {\bf %
v\cdot \sigma }\right) {\bf v}-E\left( {\bf k\cdot \sigma }\right) {\bf v}%
+iE\left( {\bf k\times v}\right) \right] w_{i}}{2\sqrt{EE^{\prime }}\sqrt{%
\left( E^{\prime }+m\right) }\sqrt{\left( E+m\right) }}  \nonumber
\end{eqnarray}
where the particle velocity ${\bf v}$ is a function of time $t$ when it
moves in the magnetic field. As it follows from general considerations, the
decay rate depends on the external magnetic field only through the
combination \cite{Ritus} 
\begin{equation}
\chi =\frac{p_{_{\perp }}}{m}\frac{B}{B_{0}}  \label{chi}
\end{equation}
Here $B_{0}=m^{2}/e$ is the critical Schwinger field and $p_{_{\perp }}\ $is
the particle momentum component orthogonal to the external magnetic field.
In the ultrarelativistic case under consideration one can write 
\begin{equation}
\chi =\frac{E}{m}\frac{B}{B_{0}}\sin \theta  \label{pz}
\end{equation}
where $\theta $ is the angle between the magnetic field ${\bf B}$ and the
particle momentum ${\bf p}$. The combination (\ref{chi}) is a Lorentz
invariant. Therefore we can simplify our calculation by considering a
particular reference frame where the electron moves along a circular
trajectory orthogonal to the magnetic field. The general case can then be
obtained by substituting $B\sin \theta $ instead of $B$ in the final result.
Considering a circular electron trajectory, one has 
\begin{equation}
{\bf v}_{1}={\bf v}\cos \left( \frac{\omega _{0}\tau }{2}\right) -\left( 
{\bf v\times h}\right) \sin \left( \frac{\omega _{0}\tau }{2}\right)
\label{v1}
\end{equation}
\begin{equation}
{\bf v}_{2}={\bf v}\cos \left( \frac{\omega _{0}\tau }{2}\right) +\left( 
{\bf v\times h}\right) \sin \left( \frac{\omega _{0}\tau }{2}\right)
\label{v2}
\end{equation}
with ${\bf h}$ being the unit vector along the magnetic field. As it follows
from kinematics, the emission in a given direction ${\bf k}$ is produced on
a small part of the path in which ${\bf v}$ turns, in the case of $\chi
\lesssim 1$, an angle $\sim m/E$ or, in the case of $\chi \gg 1$, an angle $%
\sim \chi ^{1/3}m/E$. This length is traversed on a time interval $\tau $ $%
\ll \omega _{0}^{-1}$. This short time interval gives the main contribution
to the integral over $d\tau $. Expansion in powers of $\omega _{0}\tau $
yields: 
\begin{equation}
{\bf k}\left( {\bf r}_{2}{\bf -r}_{1}\right) \simeq {\bf kv}\tau -\frac{{\bf %
kv}}{24}\tau ^{3}\omega _{0}^{2}  \label{kv}
\end{equation}
where ${\bf v}$ is the electron velocity at $t=0$. We can replace ${\bf kv}$
by $\omega $ in the last term in (\ref{kv}) because $\omega -{\bf kv\ll }%
\omega $. Thus we obtain

\begin{equation}
\exp \left[ i\frac E{E^{\prime }}\left( {\bf kr}_2{\bf -kr}_1-\omega \tau
\right) \right] \approx \exp \left[ i\frac E{E^{\prime }}\left( {\bf kv}\tau 
{\bf -}\omega \tau +\frac{\left( \omega ^2-{\bf k}^2\right) }{2E}\tau -\frac %
\omega {24}\tau ^3\omega _0^2\right) \right]  \label{12}
\end{equation}

We introduce now the following notation 
\begin{eqnarray}
M^{\dagger }M &\equiv &\left\langle i\left| \hat{Q}^{\mu \dagger }\left( 
\frac{\tau }{2}\right) \hat{Q}^{\nu }\left( -\frac{\tau }{2}\right) \right|
i\right\rangle \left( q_{\mu }q_{\nu }-g_{\mu \nu }q^{2}\right)  \label{13}
\\
&=&\frac{4\pi }{3}\left( q_{\mu }q_{\nu }-g_{\mu \nu }q^{2}\right) \frac{1}{2%
}\sum_{i}\sum_{f}R_{2}^{*\mu }R_{1}^{\nu }  \nonumber
\end{eqnarray}
Summation over the final electron polarization and averaging over the
initial electron (which we consider non-polarized) can be performed with the
help of polarization density matrices for the initial and final electron.
Keeping terms with an accuracy of $m^{4}/E^{4}$ we obtain: 
\begin{eqnarray}
M^{\dagger }M &=&\frac{4\pi }{3}\left( C_{V}^{2}+C_{A}^{2}\right) \frac{E^{2}%
}{\left( E^{\prime }\right) ^{2}}\left[ \left( \omega -{\bf kv}\right)
^{2}-\right.  \nonumber \\
&&\left. -\left( \omega ^{2}-k^{2}\right) \left( \frac{E^{2}+\left(
E^{\prime }\right) ^{2}}{2E^{2}}\frac{m^{2}}{E^{2}}+\frac{E^{\prime }}{E^{2}}%
\left( \omega -{\bf kv}\right) +\frac{\left( \omega ^{2}-k^{2}\right) }{%
4E^{2}}-\frac{m^{2}\omega ^{2}}{2E^{4}}\right) \right] +  \nonumber \\
&&+\frac{4\pi }{3}\left( C_{V}^{2}+C_{A}^{2}\right) \frac{E^{2}}{\left(
E^{\prime }\right) ^{2}}\{[{\bf kv}\left( \omega -{\bf kv}\right) -\left( 
{\bf k}\left( {\bf v\times h}\right) \right) ^{2}-  \nonumber \\
&&-\left( \omega ^{2}-k^{2}\right) \left( \left( \frac{E^{2}+\left(
E^{\prime }\right) ^{2}}{E^{2}}\right) +\frac{E^{\prime }}{E^{2}}\frac{1}{2}%
{\bf kv}\right) ]\frac{\tau ^{2}\omega _{0}^{2}}{4}+\omega ^{2}\frac{\tau
^{4}\omega _{0}^{4}}{64}\}+  \nonumber \\
&&+\frac{4\pi }{3}C_{A}^{2}\frac{m^{2}}{E^{\prime 2}}\left[ 3\frac{E^{\prime
}}{E}\left( \omega ^{2}-k^{2}\right) -\omega ^{2}\frac{\tau ^{2}\omega
_{0}^{2}}{2}\right] +  \nonumber \\
&&-\frac{4i\pi }{3}C_{V}C_{A}\frac{1}{\left( E^{\prime }\right) ^{2}}\left(
E^{\prime }+m\right) \left( \omega ^{2}-k^{2}\right) \left( {\bf kh}\right)
\tau \omega _{0}  \label{16}
\end{eqnarray}

Substitution of Eqs.(\ref{12}) and (\ref{16}) into Eq.(\ref{5}) yields 
\begin{eqnarray}
d\Gamma &=&\frac{G_{F}^{2}}{2}\frac{d\omega }{\left( 2\pi \right) ^{6}}%
\int_{0}^{\omega }dkk^{2}d\Omega  \label{17} \\
&&\times \int_{-\infty }^{\infty }d\tau \exp \left[ -i\frac{E}{E^{\prime }}%
\left( \left( \omega -{\bf kv}+\frac{\left( \omega ^{2}-{\bf k}^{2}\right) }{%
2E}\right) \tau +\frac{\omega }{24}\tau ^{3}\omega _{0}^{2}\right) \right]
M^{\dagger }M  \nonumber
\end{eqnarray}
The last term in Eq.(\ref{16}) contributes only to the azimuthal
distribution of radiated neutrino pairs, and will vanish after integration
over all directions of ${\bf k}$ with respect to the initial electron
velocity : $d\Omega =\sin \left( \vartheta \right) d\vartheta d\varphi $.
Integration over $d\varphi $ is trivial. In order to perform the next
integrations let us use instead of $\omega $ the new variable

\begin{equation}
s\equiv \frac 1{\gamma ^2}\left( \frac{E\omega }{E^{\prime }\omega _0}%
\right) ^{2/3}  \label{s}
\end{equation}
where $\gamma =m/E$ is the electron Lorentz factor, so that 
\begin{equation}
\frac \omega E=\frac{\chi s^{3/2}}{\left( 1+\chi s^{3/2}\right) }  \label{21}
\end{equation}
and introduce the following changes of variable :

\begin{equation}
z\equiv \frac{2}{\omega }\gamma ^{2}s\left( \omega -kv\cos \vartheta -\frac{%
\omega ^{2}-k^{2}}{2E}\right)  \label{z}
\end{equation}
\begin{equation}
\frac{u^{3}}{3}\equiv \frac{E}{E^{\prime }}\frac{\omega \omega _{0}^{2}}{24}%
\tau ^{3}  \label{u}
\end{equation}
\begin{equation}
a\equiv \frac{2}{\omega }\left( \omega -kv-\frac{\omega ^{2}-k^{2}}{2E}%
\right) \gamma ^{2}s  \label{a}
\end{equation}
Taking also into account the following identity for the Airy function and
its $n$-th derivatives 
\begin{equation}
\int_{-\infty }^{\infty }du\cdot u^{n}\exp \left[ -i\left( xu+\frac{u^{3}}{3}%
\right) \right] =2\pi \left( i\right) ^{n}%
\mathop{\rm Ai}%
^{\left( n\right) }\left( x\right)  \label{Airy}
\end{equation}
one obtains, with leading accuracy $1/\gamma ^{6}$

\begin{eqnarray}
&&\frac{d\Gamma }{ds}=\frac{G_F^2m^4}{16\left( 2\pi \right) ^3}\frac m\gamma 
\frac{\chi ^5s^{3+1/2}}{\left( 1+\chi s^{3/2}\right) ^4}\int_s^\infty
da\int_a^\infty dz\times  \label{spectr} \\
&&\{\left( C_V^2+C_A^2\right) \left[ 2\frac{\chi ^2s^3}{\left( 1+\chi
s^{3/2}\right) }\left( a-s\right) \left( s+z-a\right) +z^2-4s\left(
a-s\right) \right] 
\mathop{\rm Ai}%
\left( z\right)  \nonumber \\
&&+\left( C_V^2+C_A^2\right) \left\{ \left[ \left( 6+2\chi s^{3/2}-2\frac{%
\chi s^{3/2}}{\left( 1+\chi s^{3/2}\right) }\right) \left( a-s\right) 
\mathop{\rm Ai}%
^{\left( 2\right) }\left( z\right) -2s%
\mathop{\rm Ai}%
^{\left( 2\right) }\left( z\right) \right] \right.  \nonumber \\
&&\left. +%
\mathop{\rm Ai}%
^{\left( 4\right) }\left( z\right) \}+4C_A^2\left[ 3s\left( a-s\right) 
\mathop{\rm Ai}%
\left( z\right) +2s%
\mathop{\rm Ai}%
^{\left( 2\right) }\left( z\right) \right] \right\}  \nonumber
\end{eqnarray}
We have replaced the upper limit of integration over $dz$ by infinity,
because the actual limit is given by 
\begin{equation}
\frac 2\omega \left( \omega +kv-\frac{\omega ^2-k^2}{2E}\right) \gamma
^2s\simeq 2\left( 2-\frac{\omega ^2-k^2}{2E\omega }\right) \gamma ^2s\simeq
4\gamma ^2s\gg 1  \label{limit}
\end{equation}
(see Eq.(\ref{s})). In this upper limit, the Airy function tends to zero
exponentially. Performing integration by parts we get 
\begin{eqnarray}
\frac{d\Gamma }{ds} &=&\frac{G_F^2m^4}{16\left( 2\pi \right) ^3}\frac m\gamma
\frac{\chi ^5s^{3+1/2}}{\left( 1+\chi s^{3/2}\right) ^4}\times  \label{19} \\
&&\left\{ \left( C_V^2+C_A^2\right) \left[ \frac{\chi ^2s^3}{\left( 1+\chi
s^{3/2}\right) }\int_s^\infty \left[ 2+\frac 13\left( 2s+a\right) \left(
a-s\right) ^2\right] 
\mathop{\rm Ai}%
\left( a\right) da\right. \right.  \nonumber \\
&&\left. +\int_s^\infty \left[ 6+\left( a-s\right) \left( s^2+\left(
s-a\right) ^2\right) \right] 
\mathop{\rm Ai}%
\left( a\right) da-s%
\mathop{\rm Ai}%
\left( s\right) \right]  \nonumber \\
&&\left. +C_A^28s\left[ \frac 34\left( \int_s^\infty \left( s-a\right) ^2%
\mathop{\rm Ai}%
\left( a\right) da\right) +%
\mathop{\rm Ai}%
\left( s\right) \right] \right\}  \nonumber
\end{eqnarray}

Since $s$, defined by Eq.(\ref{s}), is a function of $\omega $, this formula
represents the energy spectrum of radiated neutrino pairs.

In Fig. 2 we have plotted the neutrino pair energy spectrum, as a function
of the variable $s$, obtained from Eq.(\ref{19}) (normalized so that the
integral over $s$ is equal to unity in all cases), for three values of the
characteristic parameter $\chi $. The energy distribution has a maximum for $%
s\sim 1$ if $\chi \ll 1$. Then, in this limit, radiated neutrinos and
antineutrinos have energies $\omega \sim \omega _{0}\gamma ^{3}$ , while for 
$\chi \gg 1$ the energy distribution peaks at small $s\sim \chi ^{-2/3}$.
The energy of the emitted neutrino pair, in this case, goes up to $\omega
\sim E$ , while the energy of the final particle is as small as $mB_{0}/B$.
By this reason, in the ultrarelativistic limit under consideration, we have
to require $B\ll B_{0}$. To obtain the total decay width, Eq.(\ref{19}) must
be integrated with respect to $s$ from $0$ to $\gamma /\chi $, as follows
from Eq.(\ref{21}) when $\omega $ varies from $0$ to $E-m\simeq E$. 
\begin{eqnarray}
\Gamma &=&\frac{G_{F}^{2}m^{4}}{16\left( 2\pi \right) ^{3}}\frac{m}{\gamma }%
\int_{0}^{\gamma /\chi }ds\frac{\chi ^{5}s^{3+1/2}}{\left( 1+\chi
s^{3/2}\right) ^{4}}\times  \label{total} \\
&&\left\{ \left( C_{V}^{2}+C_{A}^{2}\right) \left[ \frac{\chi ^{2}s^{3}}{%
\left( 1+\chi s^{3/2}\right) }\int_{s}^{\infty }\left[ 2+\frac{1}{3}\left(
2s+a\right) \left( a-s\right) ^{2}\right] 
\mathop{\rm Ai}%
\left( a\right) da\right. \right.  \nonumber \\
&&\left. +\int_{s}^{\infty }\left[ 6+\left( a-s\right) \left( s^{2}+\left(
s-a\right) ^{2}\right) \right] 
\mathop{\rm Ai}%
\left( a\right) da-s%
\mathop{\rm Ai}%
\left( s\right) \right]  \nonumber \\
&&\left. +C_{A}^{2}8s\left[ \frac{3}{4}\left( \int_{s}^{\infty }\left(
s-a\right) ^{2}%
\mathop{\rm Ai}%
\left( a\right) da\right) +%
\mathop{\rm Ai}%
\left( s\right) \right] \right\}  \nonumber
\end{eqnarray}
Since we assume $\gamma /\chi =$ $B_{0}/B\gg 1$, integration can be extended
up to infinity in the latter equation.

The total width can be calculated analytically in two limiting cases of
small and large value of $\chi $ , by taking into account the above
discussion. When $\chi \ll 1$ we obtain : 
\begin{equation}
\Gamma \left( \chi \ll 1\right) =\frac{49}{27}\sqrt{3}\frac{\left(
C_{V}^{2}+C_{A}^{2}\right) G_{F}^{2}}{16\left( 2\pi \right) ^{3}}\frac{m^{6}%
}{E}\chi ^{5}+\frac{7}{4}\sqrt{3}\frac{C_{A}^{2}G_{F}^{2}}{2\left( 2\pi
\right) ^{3}}\frac{m^{6}}{E}\chi ^{5}  \label{2.1}
\end{equation}

This result coincides with that given by Baier and Katkov \cite{Baier66} for 
$\nu \bar{\nu}$ synchrotron radiation of an electron, if we take $%
C_{V}=C_{A}=1$ . In the opposite case of extremely large $\ln \chi \gg 1$,
we get the leading term 
\begin{equation}
\Gamma \left( \chi \gg 1\right) =\frac{\left( C_{V}^{2}+C_{A}^{2}\right)
G_{F}^{2}}{27\left( 2\pi \right) ^{3}}\frac{m^{6}}{E}\chi ^{2}\ln \chi
\label{2.2}
\end{equation}
which also coincides with that obtained by Baier and Katkov \cite{Baier66}
(see also \cite{Borisov} and \cite{Ternov} )

\section{Single-particle approximation}

Consider now the neutrino pair emissivity due to synchrotron radiation from
a magnetized ultrarelativistic degenerate electron gas. Here some remarks
are in order. When calculating the rate of $\nu \overline{\nu }$ synchrotron
emission by an ultrarelativistic electron in vacuum we required $B\ll
B_{0}\, $in order to the final electron still be ultrarelativistic. Now,
when we consider a degenerate electron gas, one can omit this limitation
because the radiated pair energy is of the order of the medium temperature
and, consequently, it is much smaller than the energy of the initial
electron, which comes from the vicinity of the Fermi surface. To use the
above developed method in this case, one requires only two conditions: the
electron must be ultrarelativistic, i.e. $E_{F}\gg m$ , and its motion must
be semi-classical, i.e. $\omega _{0}\ll E_{F}$. The latter condition is
equivalent to $B\ll B_{0}E_{F}^{2}/m^{2}$.

As we will show, in this scenario calculations are essentially simpler than
those for vacuum, because in the matrix elements one can neglect all terms
which are of the order of, or smaller than, $T/E_{F}$. Then the contribution
of the contact weak interaction can be simply obtained from Eq.(\ref{19}) if
one takes into account that, under the conditions we considered, $\chi
s^{3/2}\sim T/E_{F}\ll 1$. Neglecting these small terms one obtains 
\begin{eqnarray}
\frac{d\Gamma }{ds} &=&\frac{G_{F}^{2}m^{4}}{16\left( 2\pi \right) ^{3}}%
\frac{m}{\gamma }\chi ^{5}s^{7/2}\left\{ \left( C_{V}^{2}+C_{A}^{2}\right)
\left[ \int_{s}^{\infty }\left[ 6+\left( a-s\right) \left( s^{2}+\left(
s-a\right) ^{2}\right) \right] 
\mathop{\rm Ai}%
\left( a\right) da\right. \right.  \label{19d} \\
&&\left. \left. -s%
\mathop{\rm Ai}%
\left( s\right) \right] +C_{A}^{2}8s\left[ \int_{s}^{\infty }\left(
s-a\right) ^{2}%
\mathop{\rm Ai}%
\left( a\right) da+%
\mathop{\rm Ai}%
\left( s\right) \right] \right\}  \nonumber
\end{eqnarray}
In the actual magnetic field of a neutron star one has, moreover, 
\begin{equation}
B\gg \frac{mT}{E_{F}^{2}}B_{0}  \label{c4}
\end{equation}
Therefore 
\begin{equation}
s\equiv \frac{m^{2}}{E_{F}^{2}}\left( \frac{\omega }{\omega _{0}}\right)
^{2/3}\sim \left( \frac{mT}{E_{F}^{2}}\frac{B_{0}}{B}\right) ^{2/3}\ll 1
\label{2.3}
\end{equation}
and one can substitute by zero the lower limit of integration in (\ref{19d})
and neglect $s$ inside the integrals. Then one can perform this integral
analytically. Keeping only leading terms and making the substitution (\ref
{pz}), we obtain the rate of neutrino pair emission due to the contact weak
interaction 
\begin{equation}
\frac{d\Gamma _{w}}{d\omega }=\left( C_{V}^{2}+C_{A}^{2}\right) \frac{%
G_{F}^{2}m^{4}}{9\left( 2\pi \right) ^{3}}n\left( E\right) \left[ 1-n\left(
E-\omega \right) \right] \frac{B^{2}}{B_{0}^{2}}\frac{\omega ^{2}}{E_{F}^{2}}%
\sin ^{2}\theta  \label{gw}
\end{equation}
Here we introduced the electron statistical factors $n\left( 1-n\right) $
with 
\[
n\left( E\right) \simeq \frac{1}{e^{\left( E-E_{F}\right) /T}+1} 
\]
taking into account the Pauli principle.

\section{In-medium effective weak interaction}

As it follows from the second Feynman diagram, the effective interaction of
neutrinos with the photon field $A_{\mu }$ arises from the medium
polarization coupling to the weak neutrino current $\bar{\nu}\gamma _{\mu
}\left( 1-\gamma _{5}\right) \nu $. Since one can neglect the magnetic field
contribution to the medium polarization, the effective vertex of this
interaction reads (see also \cite{Adams},\cite{Braaten} ) 
\begin{equation}
\frac{G_{F}}{\sqrt{2}}\Gamma ^{\mu \nu }=\frac{G_{F}}{\sqrt{2}}\frac{1}{%
\sqrt{e^{2}}}(C_{V}\Pi _{l}\left( q\right) e^{\mu }e^{\nu }+g^{\mu
i}[C_{V}\Pi _{t}\left( q\right) \left( \delta ^{ij}-n^{i}n^{j}\right)
+iC_{A}\Pi _{A}\left( q\right) \epsilon ^{ijm}n^{m}]g^{j\nu })  \label{2.6}
\end{equation}
with the notations ${\bf n=k}/k$ , $k=|{\bf k}|$ and 
\begin{equation}
e^{\mu }=\left( 1,\frac{\omega }{k}{\bf n}\right)  \label{2.7}
\end{equation}
$\Pi _{l}\left( q\right) $ and $\Pi _{t}\left( q\right) $ are the
longitudinal and transversal electromagnetic polarization functions,
respectively, and $\Pi _{A}\left( q\right) $ is the axial polarization
function . These functions, calculated in the one-loop approximation for an
ultrarelativistic degenerate electron gas are \cite{Braaten}: 
\begin{equation}
\Pi _{l}\left( q\right) =3\omega _{p}^{2}\frac{\omega ^{2}}{k^{2}}\left( 
\frac{\omega }{2k}\ln \frac{\omega +k}{\omega -k}-1\right)  \label{2.8}
\end{equation}
\begin{equation}
\Pi _{t}\left( q\right) =\frac{3}{2}\omega _{p}^{2}\frac{\omega ^{2}}{k^{2}}%
\left( 1-\frac{\omega ^{2}-k^{2}}{\omega ^{2}}\frac{\omega }{2k}\ln \frac{%
\omega +k}{\omega -k}\right)  \label{2.9}
\end{equation}
\begin{equation}
\Pi _{A}\left( q\right) =\frac{3}{2}\frac{\omega _{p}^{2}}{E_{F}}\frac{%
\omega ^{2}-k^{2}}{k}\left( \frac{\omega }{2k}\ln \frac{\omega +k}{\omega -k}%
-1\right)  \label{2.10}
\end{equation}
where $\omega _{p}$ is the plasma frequency. One can see that, when $%
q^{2}/\omega _{p}^{2}\ll 1$ , the longitudinal and transversal polarization
functions behave as : 
\begin{equation}
\Pi _{l}\left( q^{2}\rightarrow 0\right) =\frac{3}{2}\omega _{p}^{2}\ln 
\frac{4\omega ^{2}}{q^{2}}  \label{2.11}
\end{equation}
\begin{equation}
\Pi _{t}\left( q^{2}\rightarrow 0\right) =\frac{3}{2}\omega _{p}^{2}
\label{2.12}
\end{equation}
while the axial polarization function goes to zero 
\begin{equation}
\Pi _{A}\left( q^{2}\rightarrow 0\right) =\frac{3}{4}\frac{\omega _{p}^{2}}{%
E_{F}}\frac{q^{2}}{\omega }\ln \frac{4\omega ^{2}}{q^{2}}\rightarrow 0
\label{2.14}
\end{equation}
and, by this reason, it can be neglected. So, instead of (\ref{9}) one
should write 
\begin{equation}
{\cal R}^{\mu }\left( t\right) =R^{\mu }\left( t\right) +r^{\mu }\left(
t\right)  \label{2.15}
\end{equation}
where $R^{\mu }\left( t\right) $ is given by Eq. (\ref{7}), while 
\begin{equation}
r^{\mu }\left( t\right) =\frac{u_{f}^{\dagger }\left( p^{\prime }\right) }{%
\left( 2E^{\prime }\right) ^{1/2}}\gamma ^{0}\gamma ^{\lambda }\frac{%
u_{i}\left( p\right) }{\left( 2E\right) ^{1/2}}D_{\lambda \rho }\left(
q\right) \Gamma ^{\rho \mu }\left( q\right)  \label{r}
\end{equation}
is the matrix element corresponding to the second diagram. Here $D_{\lambda
\rho }\left( q\right) $ is the effective propagator for the electromagnetic
field in the medium, which in the $A_{0}=0$ gauge reads 
\begin{equation}
D_{\lambda \rho }\left( q\right) =g_{i\lambda }[-\frac{1}{\omega ^{2}-\Pi
_{l}}n^{i}n^{j}-\frac{1}{q^{2}-\Pi _{t}}\left( \delta
^{ij}-n^{i}n^{j}\right) ]g_{j\rho }  \label{2.17}
\end{equation}
As it follows from Eq.( \ref{17}) the rate of neutrino pair emission is 
\begin{eqnarray}
d\Gamma &=&\frac{G_{F}^{2}}{2}\frac{d\omega }{\left( 2\pi \right) ^{6}}%
n\left( E\right) \left[ 1-n\left( E-\omega \right) \right] \int_{0}^{\omega
}dkk^{2}d\Omega  \label{2.18} \\
&&\times \int_{-\infty }^{\infty }d\tau \exp \left[ -i\left( \left( \omega -%
{\bf kv}\right) \tau +\frac{\omega }{24}\tau ^{3}\omega _{0}^{2}\right)
\right] \left( M+\mu \right) ^{\dagger }\left( M+\mu \right)  \nonumber
\end{eqnarray}
We have introduced the statistical factors $n\left( E\right) \left[
1-n\left( E-\omega \right) \right] $ , as in previous section, and neglected
the third term in the exponential (see Eq.(\ref{12})) because $\omega \sim
T\ll E$. As it follows from (\ref{2.18}), one has to evaluate the
contribution of three terms: the first one is the contribution from contact
weak interaction, defined by 
\begin{equation}
M^{\dagger }M=\frac{4\pi }{3}\frac{1}{2}\sum_{i}\sum_{f}{\sf 
\mathop{\rm Tr}%
}\left( R_{2}^{*\mu }R_{1}^{\nu }\right) \left( q_{\mu }q_{\nu }-g_{\mu \nu
}q^{2}\right)  \label{2.19}
\end{equation}
The second one is the contribution of the effective interaction via the
intermediate virtual photon 
\begin{equation}
\mu ^{\dagger }\mu =\frac{4\pi }{3}\frac{1}{2}\sum_{i}\sum_{f}{\sf 
\mathop{\rm Tr}%
}\left( r_{2}^{*\mu }r_{1}^{\nu }\right) \left( q_{\mu }q_{\nu }-g_{\mu \nu
}q^{2}\right)  \label{mumu}
\end{equation}
and the third one is the interferential term, given by 
\begin{equation}
M^{\dagger }\mu +\mu ^{\dagger }M=\frac{4\pi }{3}\frac{1}{2}\sum_{i}\sum_{f}%
{\sf 
\mathop{\rm Tr}%
}\left( R_{2}^{*\mu }r_{1}^{\nu }+r_{2}^{*\mu }R_{1}^{\nu }\right) \left(
q_{\mu }q_{\nu }-g_{\mu \nu }q^{2}\right)  \label{2.20}
\end{equation}

The first contribution has been already evaluated, and one has now to
calculate the rate of neutrino pair emission due to weak interactions, via
the intermediate virtual photon, and the interferential term.

\section{Collective effects}

In accordance to (\ref{r}) one has

\begin{equation}
r^\mu =-C_V\xi ^i[\frac \omega k\frac{\Pi _l}{\omega ^2-\Pi _l}n^ie^\mu -%
\frac{\Pi _t}{q^2-\Pi _t}\left( \delta ^{ij}-n^in^j\right) g^{j\mu }]
\label{3.1}
\end{equation}
We introduced here a short notation for the following matrix element 
\begin{equation}
{\bf \xi }=\frac{u_f^{*}\left( p^{\prime }\right) }{\left( 2E^{\prime
}\right) ^{1/2}}{\bf \alpha }\frac{u_i\left( p\right) }{\left( 2E\right)
^{1/2}}  \label{ksi}
\end{equation}
Taking into account $r^\mu q_\mu =0$, contraction in (\ref{mumu}) yields 
\begin{eqnarray}
&&\frac{4\pi }3r_2^{*\mu }r_1^\nu \left( q_\mu q_\nu -g_{\mu \nu }q^2\right)
\label{ksiksi} \\
&=&\frac{4\pi }3C_V^2q^2\left[ \frac{\omega ^2q^2}{k^4}\left| \frac{\Pi _l}{%
\omega ^2-\Pi _l}\right| ^2\left( {\bf \xi }_2^{*}{\bf n}\right) \left( {\bf %
\xi }_1{\bf n}\right) +\left| \frac{\Pi _{tr}}{q^2-\Pi _{tr}}\right|
^2\left( \left( {\bf \xi }_2^{*}{\bf \xi }_1\right) -\left( {\bf \xi }_2^{*}%
{\bf n}\right) \left( {\bf \xi }_1{\bf n}\right) \right) \right]  \nonumber
\end{eqnarray}
Since in the case of a degenerate electron gas, the energy of the radiated
neutrino pair $\omega \sim T$ is much smaller than the initial electron
energy $E\simeq E_F$, one can neglect in the matrix elements (\ref{ksi}) all
terms which are of the order or smaller than $mT/E_F^2$. Then one obtains 
\begin{equation}
{\bf \xi }=\left( w_f^{*}w_i\right) {\bf v}\left( t\right)  \label{3.2}
\end{equation}
Summation in (\ref{ksiksi}) over the final electron polarization and
averaging over the initial electron yields 
\begin{equation}
\mu ^{\dagger }\mu =\frac{4\pi }3C_V^2q^2\left[ \frac{\omega ^2q^2}{k^4}%
\left| \frac{\Pi _l}{\omega ^2-\Pi _l}\right| ^2\left( {\bf v}_2{\bf n}%
\right) \left( {\bf v}_1{\bf n}\right) +\left| \frac{\Pi _{tr}}{\omega
^2-\Pi _{tr}}\right| ^2\left[ {\bf v}_2{\bf v}_1-\left( {\bf v}_2{\bf n}%
\right) \left( {\bf v}_1{\bf n}\right) \right] \right]  \label{3.3}
\end{equation}
Expanding ${\bf v}_2$ and ${\bf v}_1$ in the small parameter $\omega
_0^2\tau ^2$ we obtain, according to Eqs. (\ref{v1},\ref{v2}) 
\begin{equation}
\mu ^{\dagger }\mu =\frac{4\pi }3C_V^2q^2\left[ \frac{q^2}{\omega ^2}\left| 
\frac{\Pi _l}{\omega ^2-\Pi _l}\right| ^2\left( {\bf nv}\right) ^2+\left| 
\frac{\Pi _{tr}}{q^2-\Pi _{tr}}\right| ^2\left( {\bf v}^2-\left( {\bf nv}%
\right) ^2-{\bf v}^2\frac{\omega _0^2\tau ^2}4\right) \right]  \label{3.4}
\end{equation}
In the fist term in the square brackets we keep only $\left( {\bf nv}\right)
^2\sim 1$ because this term is multiplied by $q^2/\omega ^2\sim 1/\gamma ^2$
.

We introduce a special notation $\Gamma _{em}$ for the second diagram
contribution to the rate of reaction in order to distinguish it from $\Gamma
_w$. Performing integrations in

\begin{eqnarray}
\frac{d\Gamma _{em}}{d\omega } &=&\frac{G_{F}^{2}}{2}\frac{1}{\left( 2\pi
\right) ^{6}}n\left( E\right) \left[ 1-n\left( E-\omega \right) \right]
\times  \label{3.5} \\
&&\times \int_{0}^{\omega }dkk^{2}\int d\Omega \int_{-\infty }^{\infty
}d\tau \exp \left[ -i\left( \left( \omega -kvx\right) \tau +\frac{\omega }{24%
}\tau ^{3}\omega _{0}^{2}\right) \right] \mu ^{\dagger }\mu  \nonumber
\end{eqnarray}
as it was made in the previous section, we arrive to the following
expression 
\begin{eqnarray}
\frac{d\Gamma _{em}}{d\omega } &=&\frac{G_{F}^{2}C_{V}^{2}}{2}\frac{1}{%
\left( 2\pi \right) ^{4}}n\left( E\right) \left[ 1-n\left( E-\omega \right)
\right] \int_{0}^{\omega }dkk\int_{a}^{a_{1}}dz  \label{3.6} \\
&&\frac{4\pi }{3}q^{2}\left\{ \frac{q^{2}k^{2}}{\omega ^{4}}\left| \frac{\Pi
_{l}}{\omega ^{2}-\Pi _{l}}\right| ^{2}\frac{\omega ^{2}}{v^{2}k^{2}}\left(
1-\beta z\right) ^{2}%
\mathop{\rm Ai}%
\left( z\right) \right.  \nonumber \\
&&\left. +\left| \frac{\Pi _{tr}}{q^{2}-\Pi _{tr}}\right| ^{2}\left( \left(
1-\frac{\omega ^{2}}{v^{2}k^{2}}\left( 1-\beta z\right) ^{2}\right) 
\mathop{\rm Ai}%
\left( z\right) +2\beta 
\mathop{\rm Ai}%
^{(2)}\left( z\right) \right) \right\}  \nonumber
\end{eqnarray}
with the notation 
\begin{equation}
\beta \equiv \frac{1}{2}\left( \frac{\omega _{0}}{\omega }\right) ^{2/3}
\end{equation}
and where the limits of integration over $dz$ are 
\begin{equation}
a=\frac{2}{\omega }\left( \frac{\omega }{\omega _{0}}\right) ^{2/3}\left(
\omega -kv\right)  \label{3.7}
\end{equation}
\begin{equation}
a_{1}=\frac{2}{\omega }\left( \frac{\omega }{\omega _{0}}\right)
^{2/3}\left( \omega +kv\right)  \label{3.8}
\end{equation}
Since we assume 
\begin{equation}
\left( \frac{\omega }{\omega _{0}}\right) ^{2/3}=\left( \frac{\omega E_{F}}{%
eB}\right) ^{2/3}\sim \left( \frac{TE_{F}}{eB}\right) ^{2/3}\gg 1
\label{3.9}
\end{equation}
one can change the upper limit of integration over $dz$ to infinity because
the Airy function exponentially tends to zero when $z\gg 1$ . By the same
reason, the integrand is not small only if the lower limit is of the order
or less than unity. This means that only those values of $k$ which are very
close to $\omega $ contribute dominantly to the integral 
\begin{equation}
\frac{\left( \omega -k\right) }{\omega }\sim \left( \frac{\omega _{_{0}}}{%
\omega }\right) ^{2/3}\ll 1  \label{3.10}
\end{equation}
The latter inequality also means that 
\begin{equation}
q^{2}\simeq 2\omega \left( \omega -k\right) \sim \omega ^{2}\left( \frac{%
\omega _{_{0}}}{\omega }\right) ^{2/3}\ll \omega ^{2}  \label{3.11}
\end{equation}
Therefore one can expand the medium polarization function, by keeping only
leading terms for $q^{2}\rightarrow 0$ as was made in (\ref{2.11}, \ref{2.12}%
), and rewrite (\ref{3.6}) as follows 
\begin{eqnarray}
\frac{d\Gamma _{em}}{d\omega } &=&\frac{G_{F}^{2}C_{V}^{2}}{6\left( 2\pi
\right) ^{3}}n\left( E\right) \left[ 1-n\left( E-\omega \right) \right]
\omega _{0}^{2}\omega ^{2}\int_{s}^{\infty }da\int_{a}^{\infty }dz
\label{3.12} \\
&&\left[ a^{2}\left| F_{l}\left( \omega ,a\right) \right| ^{2}%
\mathop{\rm Ai}%
\left( z\right) +\left| F_{tr}\left( \omega ,a\right) \right| ^{2}\left(
\left( z-a\right) a%
\mathop{\rm Ai}%
\left( z\right) +a%
\mathop{\rm Ai}%
^{(2)}\left( z\right) \right) \right]  \nonumber
\end{eqnarray}
where 
\begin{equation}
\omega ^{2}-k^{2}\simeq \omega ^{2}\left( \frac{\omega _{_{0}}}{\omega }%
\right) ^{2/3}a  \label{3.14}
\end{equation}
was used. With good accuracy one has 
\begin{equation}
F_{tr}\left( \omega ,a\right) \equiv -\lim_{q^{2}\rightarrow 0}\frac{\Pi
_{tr}}{q^{2}-\Pi _{tr}}\simeq 1  \label{3.16}
\end{equation}
while 
\begin{equation}
F_{l}\left( \omega ,a\right) \equiv -\lim_{q^{2}\rightarrow 0}\frac{\Pi _{l}%
}{\omega ^{2}-\Pi _{l}}=-\frac{\omega _{p}^{2}\left( \ln \left( \frac{\omega 
}{\omega _{0}}\right) +\frac{3}{2}\ln \frac{4}{a}\right) }{\omega
^{2}-\omega _{p}^{2}\left( \ln \left( \frac{\omega }{\omega _{0}}\right) +%
\frac{3}{2}\ln \frac{4}{a}\right) }  \label{3.15}
\end{equation}
We focus on the condition 
\begin{equation}
\frac{T^{2}}{\omega _{p}^{2}}\ll \ln \left( \frac{T}{\omega _{0}}\right)
\label{3.17}
\end{equation}
which is typically valid for a long epoch of neutron star cooling. Then one
can also substitute $F_{l}\left( \omega ,a\right) \simeq 1$.Integration by
parts can then be performed. This yields 
\begin{equation}
\frac{d\Gamma _{em}}{d\omega }=C_{V}^{2}\frac{G_{F}^{2}m^{4}}{9\left( 2\pi
\right) ^{3}}\left( \frac{B}{B_{0}}\sin \theta \right) ^{2}n\left( E\right)
\left[ 1-n\left( E-\omega \right) \right] \frac{\omega ^{2}}{E_{F}^{2}}
\label{gem}
\end{equation}

To evaluate the interferential term we neglect in (\ref{8} ) and (\ref{9})
all terms which are proportional to $\omega $ and $k$, because $\omega \sim
T\ll E_{F}$ , and we omit all terms proportional to ${\bf \sigma }$
matrices, because $%
\mathop{\rm Tr}%
{\bf \sigma =}0$. Then instead of (\ref{8}) and (\ref{9}) we get

\begin{equation}
R^{\mu }=C_{V}\left( 1,{\bf v}\left( t\right) \right) \left(
w_{f}^{*}w_{i}\right)  \label{3.19}
\end{equation}
Taking also into account (\ref{3.1} ) and (\ref{3.2}), we obtain 
\begin{eqnarray}
\mu ^{\dagger }M+M^{\dagger }\mu &=&\frac{4\pi }{3}\frac{1}{2}%
\sum_{i}\sum_{f}\left\langle i\left| \left( {\bf R}_{2}^{*\mu }{\bf r}%
_{1}^{\nu }+\left( {\bf r}_{2}^{\mu }{\bf R}_{1}^{*\nu }\right) \right)
\right| i\right\rangle \left( q_{\mu }q_{\nu }-g_{\mu \nu }q^{2}\right) =
\label{3.20} \\
&&\frac{4\pi }{3}C_{V}q^{2}\left[ \frac{\omega }{k}\frac{\Pi _{l}}{\omega
^{2}-\Pi _{l}}\left[ \left( {\bf v}_{2}{\bf n-}1\right) \left( {\bf v}_{1}%
{\bf n}\right) +\left( {\bf v}_{1}{\bf n-}1\right) \left( {\bf v}_{2}{\bf n}%
\right) \right] \right.  \nonumber \\
&&\left. +\frac{\Pi _{t}}{q^{2}-\Pi _{t}}2\left( {\bf v}_{2}{\bf v}%
_{1}-\left( {\bf v}_{1}{\bf n}\right) \left( {\bf v}_{2}{\bf n}\right)
\right) ]\right]
\end{eqnarray}
Expansion of ${\bf v}_{2}$and ${\bf v}_{1}$ in the small parameter $\omega
_{0}^{2}\tau ^{2}$ and performing integrations analogously to those in (\ref
{3.5}) we arrive to the following expression for the interferential term: 
\begin{equation}
\frac{d\Gamma _{int}}{d\omega }=\frac{5}{6}\frac{C_{V}^{2}G_{F}^{2}}{9\left(
2\pi \right) ^{3}}\left( \frac{B}{B_{0}}\sin \theta \right) ^{2}n\left(
E\right) \left[ 1-n\left( E-\omega \right) \right] \frac{\omega ^{2}}{%
E_{F}^{2}}  \label{gint}
\end{equation}
Collecting (\ref{gw}), (\ref{gem}) and (\ref{gint}) we obtain the following
expression for the deferential rate of $\nu \overline{\nu }$ synchrotron
radiation from the degenerate ultrarelativistic electron gas of a neutron
star : 
\begin{equation}
\frac{d\Gamma }{d\omega }=\left( \frac{17}{6}C_{V}^{2}+C_{A}^{2}\right) 
\frac{G_{F}^{2}m^{4}}{9\left( 2\pi \right) ^{3}}\frac{B^{2}}{B_{0}^{2}}%
n\left( E\right) \left[ 1-n\left( E-\omega \right) \right] \frac{\omega ^{2}%
}{E_{F}^{2}}\sin ^{2}\theta  \label{3.24}
\end{equation}
Comparison of this formula with (\ref{gw}) shows that, due to collective
effects under consideration, the contribution of vector weak currents to the
rate of neutrino pair emission is almost three times more than that
calculated in the single particle approximation.

\section{Neutrino pair emissivity}

With the help of Eq.(\ref{3.24}) one can calculate neutrino pair emissivity
due to the synchrotron process in the regime we are considering

\begin{equation}
Q_{\nu }=2\int \frac{d^{3}p}{\left( 2\pi \right) ^{3}}d\omega \,\omega
\left( \frac{17}{6}\sum_{\nu }C_{V}^{2}+\sum_{\nu }C_{A}^{2}\right) \frac{%
G_{F}^{2}m^{4}}{9\left( 2\pi \right) ^{3}}\frac{B^{2}}{B_{0}^{2}}n\left(
E\right) \left[ 1-n\left( E-\omega \right) \right] \frac{\omega ^{2}}{%
E_{F}^{2}}\sin ^{2}\theta  \label{3.25}
\end{equation}
Here summation is performed over three neutrino species, see (\ref{CV}, \ref
{CA}) and the extra factor $2$ takes into account summation over the initial
electron polarization. Because of the blocking factor only electrons from a
vicinity of the Fermi surface contribute to the integral, so one can use the
following approximation 
\begin{equation}
d^{3}p=p^{2}dp\sin \theta \,d\theta \,d\varphi \simeq p_{F}E_{F}dE\sin
\theta \,d\theta \,d\varphi  \label{dp}
\end{equation}
Then integrals become trivial and we obtain 
\begin{equation}
Q_{\nu }=\frac{2\zeta \left( 5\right) }{9\pi ^{5}}\left( \frac{17}{6}%
\sum_{\nu }C_{V}^{2}+\sum_{\nu }C_{A}^{2}\right) G_{F}^{2}m^{4}T^{5}\frac{%
B^{2}}{B_{0}^{2}}  \label{3.26}
\end{equation}
A comparison with the result obtained in \cite{Kaminker1} shows that, due to
collective effects, the contribution of vector weak currents to the neutrino
pair emissivity is $17/6$ times larger than that calculated in the
single-particle approximation.

\section{Discussion and Conclusion}

We have considered collective effects in $\nu \overline{\nu }$ synchrotron
radiation from an ultrarelativistic electron gas, under conditions such the
electron motion is semiclassical. These conditions are satisfied by most
astrophysical scenarios which are of interest for this process (neutron star
crust, accretion disks and so on). We apply a calculation method which
explicitly makes use of this fact, although it takes into account the
interaction among particles in a quantum way. In order to test this method,
we evaluated the single-particle synchrotron radiation in magnetized vacuum
and compared it with previous results.

In the case of a degenerate electron gas, we performed our calculations
assuming 
\begin{equation}
\frac{mT}{E_{F}^{2}}B_{0}\ll B\ll \frac{E_{F}^{2}}{m^{2}}B_{0}  \label{c1}
\end{equation}
which typically hold in neutron star matter.

\bigskip We discussed these collective effects for the process under
consideration, which can be understood as a cloud of virtual particles and
holes, which follows the radiating electron and also produce some flux of
neutrino pairs going in the same direction as the $\nu \overline{\nu }$
radiation of the initial electron. They lead to an essential enhancement
(almost three times) of the vector weak current contribution to neutrino
pair emissivity. We have evaluated them focusing on the case 
\begin{equation}
\frac{eB}{TE_F}\ll 1  \label{c2}
\end{equation}
On one hand, this means that the temperature is high enough for many Landau
levels to be occupied above the Fermi energy. On the other hand, the same
condition means that the radiated frequency $\left( \omega \sim T\right) $
is much larger than the electron gyrofrequency and so, one can neglect the
contribution of the external magnetic field to the medium polarization
tensor. As it is obvious, if condition (\ref{c2}) is not satisfied, then
synchrotron radiation of neutrino pairs will be exponentially reduced by the
small population of excited Landau levels.

\acknowledgments
This work has been partially supported by Spanish DGICYT Grant PB94-0973 and
CICYT AEN96-1718. L.B. L. would like to thank for partial support of this
work by RFFR Grant 97-02-16501 and INTAS Grant 96-0659.

{\bf \ } {\bf \ } 

\mafigura{8cm}{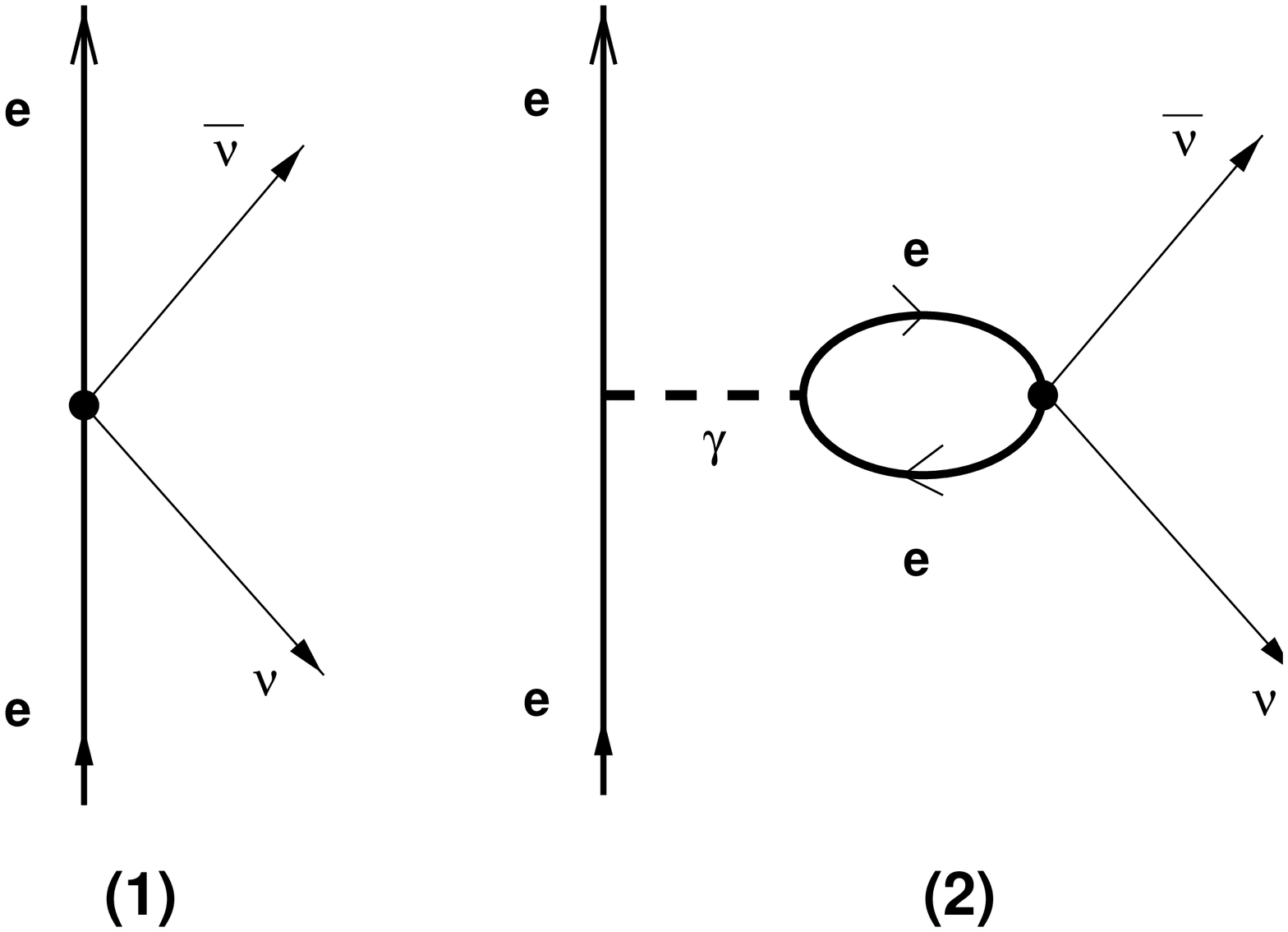}{Diagramms contributing to the matrix element of
neutrino synchrotron radiation in a dense electron plasma. The low-energy
weak interaction is shown as a filled cicle. We assume thick lines
correspond to electrons in the external magnetic field. The thick
dashed line in the second diagram represents the many-loop photon propagator
in the medium.}{Fig. 1}
\mafigura{8cm}{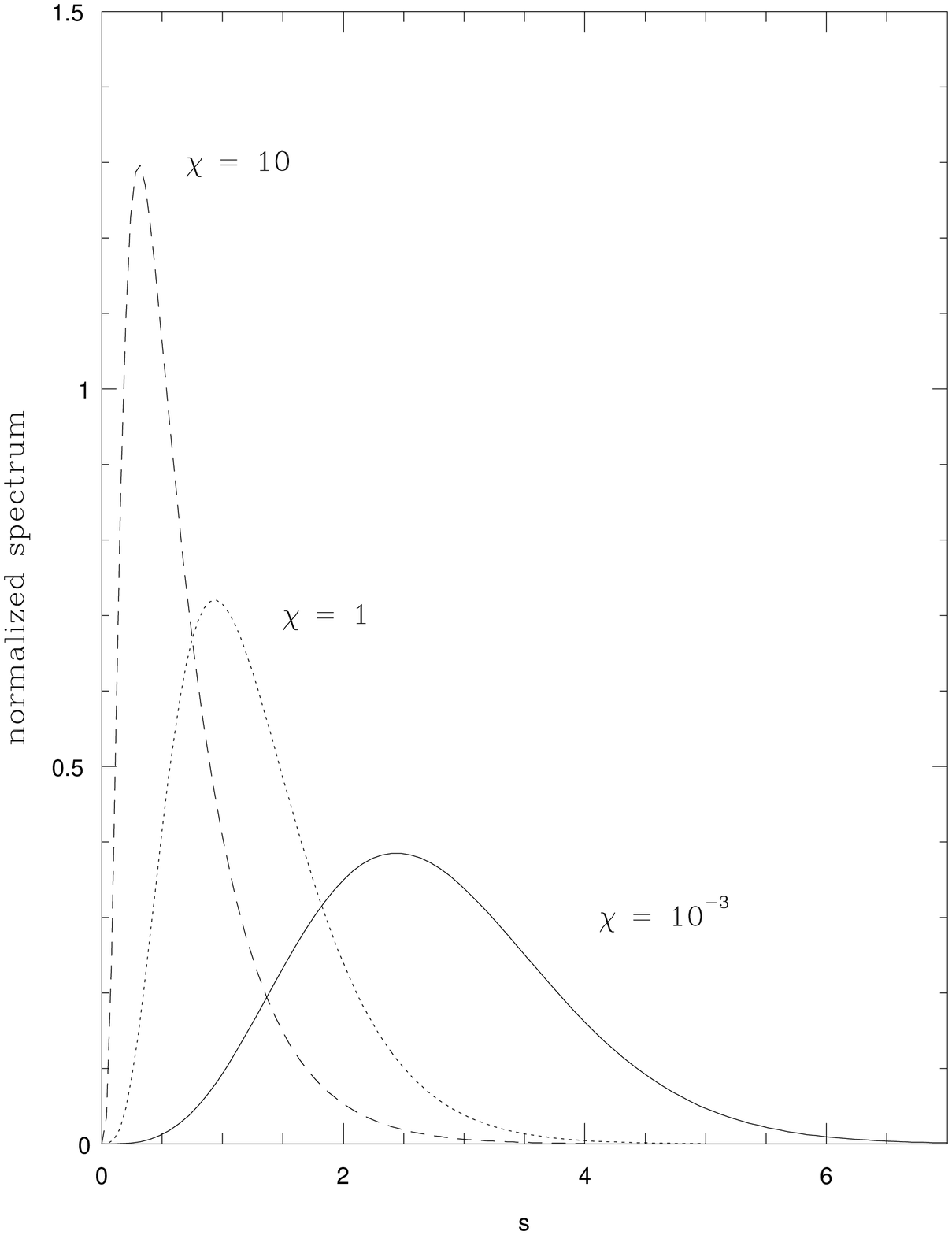}{Normalized neutrino pair energy spectrum as a
function of the variable $s$, for three values of the parameter
$\chi$, as labeled in the picture.}{Fig. 2}


\begin{references}
\bibitem{Lands}  J. D. Landstreet, Phys.Rev., {\bf 153, }1372, (1967)

\bibitem{Canuto}  V. Canuto, H. Y. Chiu and C. K. Chou, Phys.Rev., {\bf D2},
281{\bf \ } (1970)

\bibitem{YakTsch}  D. G. Yakovlev and R. Tschaepe, Astron. Nach., {\bf 302},
167 (1981)

\bibitem{Vidaurre}  A. Vidaurre, A. P\'{e}rez, H. Sivak, J. Bernab\'{e}u and
J. M${{}^{a}}$. Iba\~{n}ez, ApJ, {\bf 448, }264 (1995)

\bibitem{Kaminker1}  A. D. Kaminker, K. P. Livenfish, D. G. Yakovlev,
Soviet. Astron. Lett., {\bf 17}, 450 (1991)

\bibitem{Kaminker2}  A. D. Kaminker, K. P. Livenfish, D. G. Yakovlev, P.
Amsterdamski and P. Haensel, Phys. Rev., {\bf D46}, 3256 (1992)

\bibitem{Kaminker3}  A. D. Kaminker and D. G. Yakovlev, JETP, {\bf 76}, 229
(1993)

\bibitem{Baier}  V. N. Baier and V. M. Katkov, JETP, {\bf 52, }1422 (1967).)

\bibitem{Landau}  V. B. Berestetskii, E. M. Lifshitz and L. P. Pitaevskii,
Quantum Electrodynamics. Landau and Lifshitz Course of Theoretical Physics,
Volume 4, 2nd Edition. (Pergamon Press, Oxford, 1982), page 376.

\bibitem{Baier66}  V. N. Baier and V.M. Katkov, Sov. Phys. Doklady, {\bf 171}%
, 313 (1966).

\bibitem{Borisov}  A.V. Borisov, V.Ch. Zhukovskii and P.A. Eminov, Ivuz.
Fizika {\bf 3}, 110 (1978).

\bibitem{Ternov}  I. M. Ternov, V. N. Rodionov and A. I. Studenikin, Sov. J.
Nucl. Phys. {\bf 37, }755{\bf \ }(1983).

\bibitem{Ritus}  A. I. Nikishov and V. I. Ritus, JETP {\bf 46}, 776 (1964)

\bibitem{Adams}  J. B. Adams, M. A. Ruderman and C.-H. Woo, Phys.Rev.,{\bf %
129}, 1383 (1963)

\bibitem{Braaten}  E. Braaten and D. Segel, Phys.Rev., {\bf D48}, 1478{\bf \ 
} (1993)
\end{references}
\end{document}